\documentstyle[12pt]{article}
\begin{document}
\def\etaten{\eta_{10}}
\def\he{${\rm {}^4He\:\:}$}
\def\be{\begin{equation}}
\def\ee{\end{equation}}
\def\bc{\begin{center}}
\def\ec{\end{center}}
\def\bt{\begin{tabular}}
\def\et{\end{tabular}}
\def\mc{\multicolumn}
\def\beq{\begin{eqnarray}}
\def\eeq{\end{eqnarray}}
\def\beqd{\begin{eqnarray*}}
\def\eeqd{\end{eqnarray*}}
\def\nin{\noindent}
\def\lra{$\leftrightarrow$ }
\def\eset{$\not\!\!\:0$ }
\def\bull{$\bullet$}
\def\reaceight{$^3$He$(\alpha , \gamma )^7$Be}
\def\reacnine{$^3$H$(\alpha , \gamma )^7$Li}

\pagestyle{empty}
\rightline{{\bf CWRU-P9-94}}
\rightline{July 1994: revised Sept 1994}
\baselineskip=16pt
\vskip 0.5in
\begin{center}
\bf\large {BIG BANG NUCLEOSYNTHESIS CONSTRAINTS AND
LIGHT ELEMENT ABUNDANCE ESTIMATES }
\end{center}
\vskip0.2in
\begin{center}
Lawrence M. Krauss\footnote{Also Dept of Astronomy}
and Peter J. Kernan
\vskip .1in
 {\small\it Department of Physics\\
Case Western Reserve University\\ 10900 Euclid Ave., Cleveland, OH
44106-7079}
\vskip 0.4in
\end{center}
\centerline{{\bf Abstract}}
\noindent
To elucidate the significance of the effect of systematic uncertainties
in light element abundance estimates on cosmological bounds derivable
from Big Bang Nucleosynthesis (BBN) we present tables giving bounds on
$\Omega_{baryon}$ and $N_{\nu}$ as one changes the limits on
primordial \he and $^7$Li.  This allows us to
derive new relations between these estimates and constraints on
$\Omega_{baryon}$ and $N_{\nu}$. For example, only if the
helium mass fraction,
$Y_p \ge .245$ does $^7$Li (or D)
presently play a role in placing an upper limit on the baryon density,
and only if
$Y_p \ge .250$ does $^4$He cease to play a role in bounding $\etaten$.
All the elements combine together tend to give a stringent upper bound
of
$0.16$ on $\Omega_{baryon}$. We also find that $Y_p$ must exceed
.239 for consistency between theory and observation if
D+$^3$He/H is less than $ 10^{-4}$.
Updated nuclear reaction rates, an
updated neutron half life, Monte
Carlo techniques, and  correlations between the predicted abundances
are incorporated in our analysis.   We also discuss the
handling of systematic uncertainties in the context of statistical
analyses of BBN predictions.

\newpage
\pagestyle{plain}
\baselineskip=21pt

The theoretical analysis of BBN predictions for light element
abundances has improved greatly in recent years, allowing in principle
the derivation of very stringent constraints on various cosmological
and particle physics parameters.
 Unfortunately however, the key factor in
limiting the efficacy of this procedure is the reliability of the
inferred light element primordial abundance estimates.  Like many
quantities based on astronomical observations, these are subject to
large systematic uncertainties, many of which are difficult to
accurately estimate.

In a recent work (Kernan and Krauss 1994, hereafter KK) we underscored
the importance of considering such systematic errors when deriving BBN
constraints by demonstrating that a comprehensive analysis which used
the most up to date reaction rate uncertainties, and also incorporated
quantitatively for the first time correlations between elemental
abundances yielded, when compared with previously quoted observational
upper limits on \he, $D + ^3He$, and $^7$Li, embarassingly stringent
limits on both the number of effective neutrino types and the present
baryon density.  Indeed, it was clear that standard BBN has a very
limited range of consistency if systematic uncertainties in abundance
estimates are not allowed for. While we argued that our
results suggested the need for consideration of systematic
uncertainties, this conclusion was not as widely quoted as were the
limits we derived based on previous quoted abundance estimates which
did not explicitly incorporate such uncertainties.

Subsequently, several groups have recently assessed more carefully the
systematic uncertainties present particularly
in the primordial
\he abundance estimates  (Olive and Steigman 1994,
 Copi, Schramm and Turner 1994, Sasselov
and Goldwirth 1994), and have quoted
various new upper limits on cosmological parameters based on their
 assessments.  It is very clear, based in part on the differing
 estimates, that it is quite difficult at the present time to get an
 accurate handle on these uncertainties.

Because of this, and because we can utilize
the full statistical machinery we previously developed when comparing
predictions to ``observations", we felt it would be useful to prepare
a comprehensive table of constraints on $N_{\nu}$ and
$\Omega_{Baryon}$ for a relatively complete range of different
assumptions about light element abundances. In the
first place such a table does not exist anywhere in the
literature.  Because of the interplay between various
different elemental abundance constraints in
deriving cosmological bounds, it is not possible to easily extrapolate
previously existing limits, including our own, as abundance estimates
are varied. Different groups which advocate different
abundance limits can then only roughly translate these into
bounds on $\Omega_B$ and $N_{\nu}$. We hoped that a relatively complete
tabulation of cosmological bounds as a function of
abundance estimates would thus provide a useful reference
for researchers.  Next, the world
average neutron half life value has just been updated to be
$\tau_N=887 \pm 2 sec$ (Particle Data
Group 1994), which results in a change in all BBN constraints.  Our
new tables thus update our previous results, besides expanding upon
them.  In addition, the present analysis allows us
us to explore the role of different estimates in the
constraints, as well as the effect of correlations as the light
element abundance estimates vary.  It also allows us to
address several points which we feel are important to
consider when deriving cosmological constraints using BBN
predictions.  Finally, this analysis leads to new
 simple relations between the light element abundances and
limits on cosmological parameters such as the number of
 neutrinos\footnote{We remind the reader that $N_{\nu}$
represents the effective number of relativistic degrees of freedom in
the radiation gas during the BBN era, and is thus merely bounded
below by the actual number of light neutrino species present in
nature.}, $N_{\nu}$,  and the baryon to photon ratio
$\etaten$, defined by $\Omega_{B} =.00366 h^{-2}
(T/2.726)^3
\etaten$, where the Hubble constant is defined as $100 h$ km/sec/Mpc.

\bigskip
\noindent{\bf BBN Predictions and Observations: Systematics,
Correlations and Consistency}

The chief developments of recent years which have affected the BBN
predictions for light element abundances include: an updated BBN code,
a more accurate measured neutron half life
(Particle Data Group 1992,1994)
and the determination of BBN uncertainties via Monte Carlo analysis;
(Krauss and Romanelli 1990; Smith, Kawano and Malaney 1993).  Most
recently, we created (KK) an updated Monte
Carlo code to account both for what was then
the newest measured neutron
half life, greater numerical accuracy
(Kernan 1993) and also for new higher order effects in weak rates
(Seckel 1993). The net effect of these
changes is to both reduce the statistical error on the predicted value
of
$Y_p$, and also raise the predicted abundance by an
$\etaten$-independent factor of $+.0031$ compared to the
value used in previous published
analyses (Walker {\em et al} 1991; Krauss and Romanelli 1990).
 See KK for a more detailed
description of our analysis.

We present here an updated figure for the
predicted elemental abundances as a function of
$\etaten$ (figure 1).  However, as we
stressed in KK, this standard figure
should not be used alone to derive confidence limits on cosmological
and particle physics parameters when comparing theoretical
predictions and observations.  Because the various elemental
abundances are correlated deriving a limit using a single
element throws out valuable information from other
elements which, if incorporated, could lead to more stringent
constraints.  Stated another way, the predicted
elemental abundances are generally not statistically independent.
For example, there is a strong anti-correlation between $Y_p$ and the
remnant D +$^3$He abundance (the normalized covariance ranges from
-0.7 to -0.4 in the
$\etaten$ range of interest).   Thus, if one generates 1000
predictions using a Monte Carlo scheme, those where the predicted
\he is lower than the mean, which therefore may be allowed by some
fixed observational upper bound
$Y_p$, will also generally predict a larger than average remnant
D+$^3$He/H abundance, which in turn could exceed the
observational upper bound on this combination.  Ignoring this
correlation will result in a bound which is at the very least not
statistically consistent.  As we showed in KK, including such
correlations in our analysis had a significant effect on limits on the
number of neutrinos, and a less dramatic, but still noticable effect
on limits on $\etaten$.

Of course, if systematic uncertainties in the inferred primordial
element abundances are dominant, one might wonder whether one need
concern oneself with the proper handling of statistics in the
predicted range.  There is, after all, no well defined way to treat
systematic uncertainties statistically.  For example, should one treat
a parameter range governed by systematic uncertainties as if it were
gaussianly distributed, or uniformly distributed?  The latter
is no doubt a better approximation--i.e. a large deviation within some
range may be as equally likely as a small deviation. But how should one
handle the distribution for extreme values?  Clearly it cannot remain
uniform forever.

Thankfully, there are two factors which make the comparison
of predictions and observations less ambiguous in the case of BBN:

(1) Because the allowed range in the observationally inferred
abundances is much larger than the uncertainty in the predicted
abundances, any constraint one deduces by comparing the two depends
merely on the upper { \it or} lower observational limit for each
individual element, and not only both at the same time.  Thus, one is
not so much interested in the entire distribution of allowed abundances
as one is in one extremum of this distribution.

(2) Systematic uncertainties dominate for the
observations, while statistical uncertainties dominate for the
predictions.

Both of these factors suggest that a conservative but still well
defined approach involves setting {\it strict} upper limits on
$Y_p$, D+$^3$He, and $^7$Li, and a lower limit on D, which incorporate
the widest range of reasonably accepted systematic uncertainties.
 Determining what is reasonable in
this sense is of course where most of the ``art" lies.  We will return
to this issue shortly.  Nevertheless, once such limits are set and
treated as strict bounds, then one can compare correlated predictions
with these limits in a well defined way.  In this way one replaces the
ambiguity of properly treating the distribution of observational
estimates with what in the worst case may be a somewhat arbitrary
determination of the extreme allowed observational values.

Clearly all the power, or lack thereof, in this procedure lies in
the judicious choice of observational upper or lower limits.  Because
of our concern about the ability at present to prescribe
such limits we present below results for a variety of
them.  Nevertheless, we do wish to stress that once one
does choose such a set, it is inconsistent not to use all of it
throughout in deriving ones constraints.  If one uses one
observational upper limit for
$Y_p$, for example, to derive constraints on the number of neutrinos,
but does not use it when deriving bounds on the baryon density, then
probably one has not chosen a sufficiently conservative bound on $Y_p$
in the former analysis.   It has been argued that a weak,
logarithmic, dependence of
$Y_p$ on
$\etaten$ invalidates its use in deriving bounds on the latter
quantity.  It is one of our more interesting conclusions that not
only can this argument be somewhat misrepresentative for an
interesting range of
$Y_p$ values, but that until
$Y_p$ exceeds statistically derived upper limits by a large amount, it can
continue to play a signifcant role in bounding $\etaten$ from above.
(Note that the lower bound on $\etaten$ is presently governed
by the observational upper limit on the D+$^3$He.  For a discussion of
this bound on $\etaten$ see KK and (Krauss and Kernan 1994).

Before
proceeding to give our results, we briefly outline the rationale for
the range of limits on light element abundances we choose to explore
here.

\bigskip
\noindent{\bf Abundance Estimate Uncertainties:The Range}

It is beyond the scope of this work to
 examine in detail the observational
uncertainties associated with the determination of primordial
light element abundances.  Our purpose instead is to exploit
recent observational and theoretical estimates of these uncertainties
in order to examine how BBN constraints will be affected by
incorporating such uncertainties.  Thus we merely provide here a
very brief review of the recent literature.  The reader is referred
to the cited papers and references therein for further details.

(a) $^4$He:  By correlating $^4$He abundances with metallicity for
various heavy elements including O,N and C, in low-metallicity HII
regions one can attempt to derive a "primordial"
abundance defined as the intercept for zero metallicity.  This can be
determined by a best fit technique, assuming some linear or quadratic
correlation between elemental abundances (i.e. see Peimbert, and
Torres-Peimbert 1974;  Pagel,Terlevich and Melnick 1986; Pagel,
Simonson, Terlevich,and Kennicutt 1992;
Walker {\em et al} 1991).
The statistical errors
associated with such fits are now small.  Best fit values obtained
typically range from .228-.232, with statistical "1$\sigma$" errors on
the order of .003-.005.  This argument yields the upper limit of .24
(Walker {\em et al} 1991)
which has been oft quoted in the literature. Recently
this number has begun to drift upwards slightly.  New
observations of HII regions in metal poor galaxies have tended
to increase the statistically derived zero intercept
value of $Y_p$ by perhaps .005 (i.e. (Skillman et al 1993,
Olive and Steigman 1994)).
In addition, the recognition that systematic, and
not statistical uncertainties may dominate any
such fit has become more widespread recently. The key
systematic uncertainty which interferes with this procedure is the
uncertainty in the $^4$He abundance determined for each individual
system, based on uncertainties in modelling HII regions,
ionization, etc used to translate observed line strengths into mass
fractions.  Many observational factors come into play here
(see (Skillman et al 1994) for a discussion of observational
uncertainties), and people
have argued that one should add an extra systematic uncertainty of
anywhere from .005-.015 to the above estimate.  Clearly thus, one
should examine implications of
$^4$He abundances in the range .24-.25.  We shall show that for
$Y_p$ above .25; (a) \he becomes unimportant for bounding $\etaten$,
and (b) the effect on bounds on $N_{\nu}$ can be obtained by
straightforward extrapolation from the data obtained for the range
.24-.25.

(b) $^7$Li:  It is by now generally accepted that the primordial
abundance of $^7$Li is closer to the Spite Pop II plateau than the
Pop I plateau.  Nevertheless, even if one attempts to fit the
primordial abundance by fitting evolutionary models to the Pop II
data points (Deliyannis {\em et al } 1989), assuming no depletion,
one still finds an 2$\sigma$ upper limit as large as
 $ 2.3 \times 10^{-10}$.  The role of rotationally induced depletion is
still controversial.  It is clear some such depletion is expected,
and can be allowed for (Pinsonneault, Deliyannis and Demarque 1992),
but observations of $^6$Li,
which is more easily depleted, put limits on the amount of
$^7$Li depletion which can be allowed.  We will assume an extreme
factor of 2 depletion as allowable, and thus we
explore how cosmological bounds are affected by a $^7$Li upper
limit as large as $ \approx 5 \times 10^{-10}$.

(c) D and D+$^3$He:  We take the solar system D abundance of $ 2
\times 10^{-5}$ as a safe firm lower bound on D, and the previously
quoted upper limit of $10^{-4}$ as a firm upper limit on D+$^3$He
(Walker {\em et al} 1991).   The recent Songaila {\em et al} (1994)
 result for D (see also (Carswell {\em et al} 1994) ,
which is larger than this upper limit, is in apparent conflict with
another similar measurement, and with estimates of the pre-solar
D+$^3$He abundance, and there are preliminary
reports of contradictory data taken along other lines of sight.
  In any case, the dramatic change in BBN limits
should the former result be confirmed is discussed in great detail in
(Krauss and Kernan 1994), so we do not discuss this possibility further
 here.

\bigskip
\noindent{\bf Results and Analysis}

Tables 1-3 give our key results. The data were obtained using 1000
Monte Carlo BBN runs at each value of
$\etaten$, with nuclear reaction rate input parameters chosen as
Gaussian random variables with appropriate widths (see KK for
details) .  In each case the number of runs which resulted in
abundances which satisfied the joint constraints obtained by using
combinations of the upper limits on
$^4$He, $^7$Li, and D+$^3$He or the lower limit on D was determined.
Limits on parameters were determined by varying these until less than
50 runs out of 1000 (up to
$\sqrt{N}$ statistical fluctuations) satisfied all of the
constraints.

Table 1 displays the upper limit on  $N_{\nu}$ for various values of
$Y_p$.  As is shown, this was governed by the combination of $^4$He
and D+$^3$He upper limits.  Shown in the table are the number of
acceptable runs out of 1000 when the two elemental bounds are
considered separately and together, for an $\etaten$ range which was
 found to maximize the number of acceptable models. Throughout the
$Y_p^{max}$ region from .24 to .25, both the $Y_p$ and D+$^3$He
limits play a roughly equal role in determining the maximum value of
$N_{\nu}$.  We are able to find a remarkably good analytical fit for
the maximum value of $N_{\nu}$ as a function of $Y_p$ as follows:
\begin{equation}
N_{\nu}^{max} =3.07 + 74.07(Y_p^{max} - .240)
\end{equation}
The linearity of this relation is striking over the whole region
from .24 to .25 in spite of the interplay between the two different
limits in determining the constraint.
Note also that this relation differs from than that quoted in
Walker {\em et al} (1991)
 between $Y_p$ and $N_{\nu}$ in that the slope we find
is about
$13\%$ less steep than that quoted there.  The two formulae are not
strictly equivalent in that the one presented in
Walker {\em et al} (1991)
presented the best fit value of $Y_p$ determined in terms of
$N_{\nu}$, while the present formula gives a relation between the
maximum allowed values of these parameters, based on limits
on the { \it combination} $Y_p$ and D+$^3$He, and on the width of
the predicted distribution.  In this sense, eq. (1) is the
appropriate relation to utilize when relating bounds on $Y_p$ to
bounds on $N_{\nu}$.

Tables 2 and 3, which display the upper bounds on $\etaten$, are
perhaps even more enlightening.  They demonstrate the senitivity of
the upper limit on
$\etaten$ and hence
$\Omega_{baryon}$ to the various other elemental upper limits
as
$Y_p$ is varied.  Several features of the data are striking.  First,
note that $^4$He completely dominates in the determination of the
upper limit on
$\etaten$ until
$Y_p$ =.245, even for the most stringent chosen upper limit on
$^7$Li.  If this limit on $^7$Li is relaxed, then $^4$He dominates
as long as the upper limit on $Y_p \le$.248!  Also note
that the ``turn on" in significance of the $^7$Li contribution to
the constrain is somewhat more gradual than the ``turn off" of the
$^4$He constraint.  The former turns on over a range of $\etaten$ of
about 2, while the latter turns off over a range of about 1-1.5.
This gives one some idea of the size of the error introduced in
determining upper bounds by using only either element alone, rather
than the combination.  Next, for a
$Y_p$ upper limit which exceeds .248, the lower bound on D begins to
become important.  It quickly turns on in significance so that by the
time the upper limit on $Y_p$ is increased to .25, $^4$He essentially
no longer plays a role in bounding $\etaten$.  Finally, note that both
the relaxed bound on
$^7$Li and the D bound converge in significance at about
the same time, so that for $\etaten > 7.25$, both constraints are
significantly violated.   This implies a ``safe" upper limit on
$\etaten$ at this level, which corresponds to an upper bound
$\Omega_{baryon} \le .163$, assuming a Hubble constant in excess of
40 km/sec/Mpc.   We
again stress that a value this large is only allowed if
$Y_p$ exceeds .250.   If, for example, $Y_p \le .245$, then the
upper bound on $\Omega_{baryon}$ is essentially completely determined
by
$^4$He and is then at most 0.11.  These limits may be compared to
recent estimates of $\Omega_{baryon}$ based on
X-ray determinations of the baryon fraction in clusters
(White {\it et al} 1993).

One final comment on the role of $Y_p$ in constraining $\etaten$:
It has been stressed that because of the logarithmic dependence of the
former on the latter, that $Y_p$ cannot be effectively used to give a
reliable upper bound on $\etaten$.  This is somewhat deceptive,
however. We can compare how much more sensitive the bound
on $\etaten$ is to $Y_p$ than the bound on $N_{\nu}$ is by making a linear
fit to the former relation and comparing it to (1).  If we do this, we
find first that the linear fit is quite good out to $Y_p$ as large
as .245 (after which a quadratic fit remains good all the way
out to .248, where the D and relaxed $^7$Li bounds begin to take
over), and is given by
\begin{equation}
\eta_{10}^{max} \approx 3.22 + 354(Y_p^{max} - .240)
\end{equation}
Seen in these terms, the $\etaten$ upper limit is approximately 4.5
times more sensitive to the precise upper limit chosen for $Y_p$
than is the $N_{\nu}$ upper limit.  Thus, while there is no doubt that
varying the upper limit on $Y_p$ has a more dramatic effect on the
upper bound one might derive for $\etaten$ than it does for
constraining $N_{\nu}$,the quantitative nature of the
relative sensitivities is perhaps displayed,
 for the relevant
range of $Y_p$, by comparing the linear approximations
 presented here than by
discussing logarithmic vs linear dependencies.
  More important, even recognizing the increased sensitivity of
$\etaten$ on
$Y_p$, unless one is willing to accept the possibility of a rigid upper
bound on $Y_p$ greater than .247, it is overly conservative to ignore
$^4$He when deriving BBN bounds on $\etaten$.

Finally, we update one other quantity of importance for the
comparison of BBN predictions with observations: the
minimum value of $Y_p$ such that BBN predictions are
consistent with observation.  We explored the range
of $\etaten$ allowed at the 95$\%$ confidence level (i.e. 50 out of 1000
models) as the value of ${Y_p}^{max}$ was reduced.  For $Y_p \le .239$ no
range of $\etaten$ was allowed when this constraint was combined
with the D + $^3$He bound.  Previously we derived a lower
bound on $Y_p$ of .238 if D+$^3$He was used alone to first bound $\etaten$,
and then the $\etaten$ value was used to bound $Y_p$ (to compare
to earlier such bounds (i.e. (Krauss and Romanelli 1990)).
The new neutron half life would not change that bound.  However in any
case
the newly derived bound of .239 obtained using the correlated constraints
is more stringent, and more consistent.   If the primordial helium
abundance is determined empirically to be less than
this value with great confidence, and the D + $^3$He upper limit
remains stable, standard BBN would
be inconsistent with observation.

\bigskip
\noindent{\bf Conclusions:}
There can be little doubt that the present ability of BBN to
constrain cosmological parameters is almost completely governed by
systematic uncertainties in our inferences of the actual light element
primordial abundances.  Nevertheless, the fact that such systematic
uncertainties need not be gaussian does not block our ability to
utilize the statistically meaningful uncertainties in BBN
predictions.  As long as we are willing to quote conservative
one-sided limits on the various abundances which incorporate reasonable
estimates of the systematic uncertainties then the determination of
what confidence levels can be assigned to various theoretical
predictions is straightforward.  Moreover, as the observational limits
on various elemental abundances is varied, the significance of the
different elements for constraining cosmological parameters varies.
In addition, for a non-trivial range in $\etaten$, correlations
exist between the various abundance predictions, and a self consistent
use of all available constraints is important.  Finally,
$Y_p$, in spite of its systematic uncertainty, plays a dominant role
unless one is willing to accept an upper limit of greater than
.247.  Beyond that, the convergence of D and $^7$Li limits suggest a
safe upper bound of on the baryon density today of less than 16$\%$ of
closure density.

As time proceeds and more independent observations are made we
will undoubtedly get a better handle on the systematic
uncertainties which presently limit the efficacy of BBN
constraints.  Until then, the updated tables and relations presented here
should provide a useful reference to
 allow researchers to translate their own
limits on the
light element abundances into meaningful bounds on $N_{\nu}$ and
$\etaten$.

\bigskip

LMK thanks the Aspen Center for Physics for hospitality while much
of this work was carried out, and also thanks George Smoot for alerting
us to the likelihood that the neutron half life would be revised
in the newest Particle Data Tables compilation.

\newpage
\bc {\LARGE REFERENCES} \ec
\noindent
Carswell, R.F. et al 1994, {\em MNRAS} {\bf 268}, L1\\
\noindent
T.Copi, D.N.Schramm, M.S. Turner 1994 {\em preprint} U. Chicago\\
\noindent
Deliyannis, C.P.{\em et al} 1989 {\em Phys.Rev.Lett.} {\bf 62}, 1583 \\
\noindent
Kernan, P.J. 1993 Ph.D.Thesis Ohio State University,
 UMI-94-01290-mc \\
\noindent
Kernan, P.J., and Krauss,L.M. 1994 {\em Phys.Rev.Lett.} {\bf 72}, 3309 \\
\noindent
Krauss, L.M. and Kernan, P.J. 1994 {\em Ap.J.} in press \\
\noindent
Krauss,L.M.,  and Romanelli,P. 1990 {\em Ap.J.}, {\bf 358}, 47 \\
\noindent
Olive, K.A. and Steigman, G. 1994 {\em preprint} UMN-TH-1230/94 \\
\noindent
Pagel, B.E.J., Simonson, E.A., Terlevich, R.J., and Kennicutt Jr,
R.C. 1992, {\em Mon. Not. R. Astr. Soc.} {\bf 255}, 325
\\
\noindent
Pagel, B.E.J. Terlevich, R.J. and Melnick, J. 1986 {\em P.A.S.P.} {\bf 98},
1005 \\
\noindent
Particle Data Group 1992, {\em Phys.Rev.}{\bf
D45} S1 \\
\noindent
Particle Data Group 1994, {\em Phys.Rev.}{\bf
D50} 1173 \\
\noindent
Peimbert,M and Torres-Peimbert, S. 1974 {\em Ap J} {\bf 193}, 327\\
\noindent
Pinsonneault, M.H. , Deliyannis C.P. and Demarque, P. 1992
{\em Ap. J. Supp} {\bf 407} 699\\
\noindent
Sasselov, D.D. and  Goldwirth, D. 1994 {\em preprint} astro-ph 9407019 \\
\noindent
Seckel,D. 1993 Bartol Preprint BA-93-16 hep-ph/9305311 \\
\noindent
Skillman E.D. {\em et al} 1993 {\em Ann. New York Acad.
Sci.} {\bf 688} 739\\
\noindent
Skillman, E. D. {\em et al} 1994, {\em Ap. J.} {\bf 431} 172\\
\noindent
Smith, M.K., Kawano, L.H., and Malaney, R.A.
1993 {\em Ap.J.Supp.} {\bf 85}, 219 \\
\noindent
Songaila, A., Cowie, L.L., Hogan C.J., and Rugers, M. 1994  {\em Nature}
{\bf 368}, 599 \\
\noindent
Walker, T.P {\em et al}  1991
{\em Ap. J.} {\bf 376}, 51 \\
\noindent
White, S.D.M. {\em et al}. 1993 {\em
Nature} {\bf 366},429 \\

\bc {Table 1: \he Abundance Estimates \& $N_{\nu}$ limits}
\bt{||c||c|lccccr||}  \hline
$Y_p$& $N_{\nu_{max}}$ &\mc{6}{c||}{\#
allowed models:} \\&&
\mc{6}{c||}{\{$^4$He \& [D+$^3$He]\}($^4$He:D+$^3$He)}
\\
 \hline &&&$\eta_{10}$=2.75&2.80&2.85&2.90  & \\
.240
&3.07&&40(603:148)&52(429:254)&46(268:376)&38(170:534)& \\
 \hline &&&$\eta_{10}$=2.80&2.85&2.90&2.95  & \\
 .241
&3.14&&38(532:171)&46(354:309)&39(219:470)&35(131:625)& \\
\hline
.242
&3.21&&41(562:154)&55(451:276)&53(272:423)&52(163:616) & \\
 \hline
.243
&3.29&&17(588:110)&32(410:220)&46(266:378)&36(184:513) & \\
\hline
.244
&3.36&&30(669:102)&44(501:187)&38(353:336)&40(216:464) & \\
\hline
&&&$\eta_{10}$=2.85&2.90&2.95&3.00  & \\
.245&3.43&&50(598:173)&68(449:296)&64(308:427)&54(173:586) & \\
\hline
.247
&3.59&&27(635:84)&30(480:184)&47(338:306)&39(185:488) & \\
 \hline
&&&$\eta_{10}$=2.95&3.00&3.05&3.10  & \\
.250
&3.82&&45(491:207)&47(364:374)&50(225:495)&32(131:587) & \\
 \hline
\et \ec
\bigskip
\bc {Table 2: \he and $^7$Li Abundance Estimates \& $\eta_{10}$
limits}
\bt{||l||c|c||c|c||}  \hline
$Y_{p_{max}}$&$\eta_{10_{max}}$& \# allowed
models: &
$\eta_{10_{max}}$ &\# allowed
models:
\\
&($^7$Li$_{-10}<$
2.3)&\{$^4$He \& $^7$Li\} ($^4$He:$^7$Li)&($^7$Li$_{-10}<$ 5)&
\{$^4$He \& $^7$Li\} ($^4$He:$^7$Li) \\
\hline .240&3.26&56 (60:998)& 3.26&56 (60:1000)  \\
\hline .241&3.55&45 (45:986)& 3.55&45 (45:1000)  \\ \hline
.242&3.89&45 (47:905)& 3.89&47 (47:1000)  \\ \hline
.243&4.26&50 (60:626)& 4.27&46 (46:1000) \\ \hline
.244&4.64&48 (92:296)& 4.71&49 (49:1000) \\ \hline
.245&5.01&45 (211:118)&5.23&62 (62:984) \\ \hline
.246&5.23&51 (679:62)&5.80&46 (50:810) \\ \hline
.247&5.25&52 (997:52)&6.36&48 (80:500) \\ \hline
\et \ec

\bigskip
\bc {Table 3: $^4$He, D and $^7$Li Estimates \& $\eta_{10}$
limits ($^7$Li$_{-10}<$5;
$D_{-5}>2$)}
\bt{||l||c|c||}  \hline
$Y_{p_{max}}$&$\eta_{10_{max}}$& \# allowed
models:
\\
&&\{$^4$He \& D \& $^7$Li\} ($^4$He \& D:$^4$He \&
$^7$Li:D \& $^7$Li)
($^4$He:D:$^7$Li)\\
\hline
.248&6.94&48 (136:53:156) (178:516:203) \\ \hline
.249&7.22&52 (177:101:64) (654:217:136) \\ \hline
.250&7.24&47 (191:113:47) (995:191:113) \\ \hline
\et \ec

\newpage

\clearpage
\noindent {\bf Figure Captions}

\vskip 0.2in

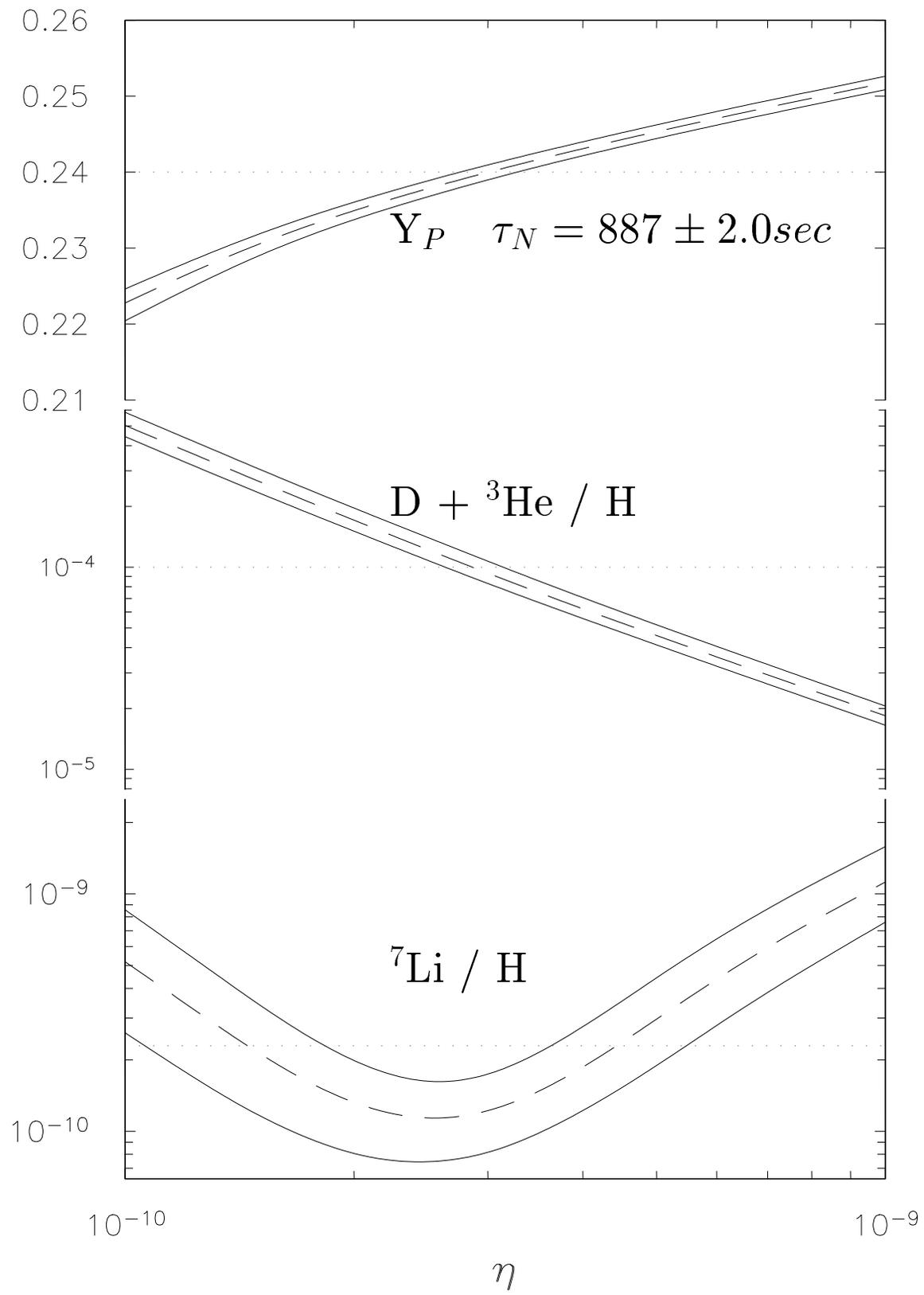
\noindent Figure 1: BBN Monte Carlo predictions as a function of
$\etaten$.  Shown are symmetric  $95 \%$ confidence limits on each elemental
abundance.

 \clearpage

\end{document}